\documentclass[twocolumn,nofootinbib,superscriptaddress]{revtex4-1}
\usepackage{latexsym,epsfig,amssymb, amsmath, nicefrac}

\usepackage[utf8]{inputenc}
\usepackage{amsmath}
\usepackage[top=1in, bottom=1in, left=1in, right=1in]{geometry}
\usepackage{color}

\newcommand{\bea}{\setlength\arraycolsep{2pt} \begin{eqnarray}}
\newcommand{\eea}{\end{eqnarray}}
\newcommand{\nn}{\nonumber}

\usepackage{hyperref}

\newsavebox{\uuunit}
\sbox{\uuunit}
{\setlength{\unitlength}{0.825em}
	\begin{picture}(0.6,0.7)
	\thinlines
	\put(0,0){\line(1,0){0.5}}
	\put(0.15,0){\line(0,1){0.7}}
	\put(0.35,0){\line(0,1){0.8}}
	\multiput(0.3,0.8)(-0.04,-0.02){12}{\rule{0.5pt}{0.5pt}}
	\end {picture}}

\def\be{\begin{equation}}
\def\ee{\end{equation}}
\def\ba{\begin{array}}
\def\ea{\end{array}}
\def\bea{\begin{eqnarray}}
\def\eea{\end{eqnarray}}
\def\bd{\begin{displaymath}}
\def\ed{\end{displaymath}}

\def\nn{\nonumber}

\def\a{\alpha}
\def\b{\beta}
\def\g{\gamma}
\def\G{\Gamma}
\def\d{\delta}

\def\f{\phi}

\def\l{\lambda}
\def\L{\Lambda}
\def\m{\mu}
\def\n{\nu}
\def\r{\rho}
\def\s{\sigma}

\def\nn{\nonumber}
\def\cD{\mathcal{D}}

\def\cL{\mathcal{L}}

\begin{document}

\vspace{1cm}

\title{Galileons as the Scalar Analogue of General Relativity}

\author{Remko Klein}
\email{remko.klein@rug.nl}
\author{Mehmet Ozkan}
\email{m.ozkan@rug.nl}
\author{Diederik Roest}
\email{d.roest@rug.nl}
\affiliation{Van Swinderen Institute for Particle Physics and Gravity, University of Groningen, \\ Nijenborgh 4, 9747 AG Groningen, The Netherlands}

\begin{abstract}
We establish a correspondence between general relativity with diffeomorphism invariance and scalar field theories with Galilean invariance: notions as the Levi-Civita connection and the Riemann tensor have a Galilean counterpart. This suggests Galilean theories as the unique non-trivial alternative to gauge theories (including general relativity). Moreover, it is shown that the requirement of first-order Palatini formalism uniquely determines the Galileon models with second-order field equations, similar to the Lovelock gravity theories. Possible extensions are discussed.

\end{abstract}	

\maketitle

\section{Introduction}

Given the ubiquity of scalar fields in many areas of physics, ranging from elementary particles and condensed matter to cosmology, one would like to understand what the most general interactions of such a field are. Already at the classical level this question turns out to have an intricate answer, both from a theoretical as well as a phenomenological viewpoint.

Scalar Lagrangians often only include the field {$\phi$  itself} and its first derivative $\partial_\m \phi$. Dependence {on higher-order derivatives} generically leads to field equations beyond second-order in derivatives, signalling additional degrees of freedom. By virtue of the Ostrogradski theorem \cite{MO},  this induces a ghost-like excitation. However, this fatal feature can be eliminated by only allowing specific combinations of the second derivatives to appear in the Lagrangians. The resulting field equations can either be purely 2$^{\rm nd}$-order or include 0$^{\rm th}$- and  1$^{\rm st}$-order terms. Such theories are referred to as (generalized) Galileons, respectively \cite{Rattazzi,Deffayet:generalized} (see also \cite{Fairlie}).

Galileons have found wide-spread applications in cosmology. For instance, they arise as the higher-order interactions in the DGP model \cite{DGP}, and when coupled to gravity allow for a self-accelerating universe without a cosmological constant \cite{Deffayet:2000uy, Deffayet:2001pu} and without ghost instabilities \cite{Rattazzi}. Moreover, Galileons obey a non-renormalization theorem \cite{NR1, NR2}, and thus can play an essential role for the construction of radiatively stable inflationary models \cite{GalInf}. Violating the Null Energy Condition, Galileons can lead to stable alternatives to inflation \cite{NEC}.  Finally, the so-called \textit{Fab-Four} subset of Galileons gives rise to self-tuning of the cosmological constant \cite{FabFour}.

These theories draw their name from the symmetry
\begin{align}
\phi \rightarrow \phi+ c + v_\mu x^\mu \,. \label{Gal}
\end{align}
This arises from a particular non-relativistic contraction of Poincar\'{e}$_{5}$ in which $\phi$ denotes the transverse position of a three-brane \cite{Gomis:2000, Gomis:2004}. In contrast, the usual Galilean symmetry $x^i \rightarrow x^i + c^i +v^i t $ arises from a non-relativistic contraction of Poincar\'{e}$_{4}$, with $x^i$ denoting a point particle position in the spatial dimension.

Purely second-order field equations are manifestly invariant under this Galilean symmetry. However, the set of theories for which this holds (which we will refer to as Galilean theories) is much larger, as it allows for any higher-order term.  We will demonstrate that it is actually the combination of Galilean symmetry plus the requirement of an underlying first-order formalism that leads to the second-order Galileons of \cite{Rattazzi}. This is reminiscent of general relativity, in which the set of diffeomorphism invariant theories generically leads to higher-order field equations. Only the subset of Lovelock gravities allows for an underlying first-order, Palatini formalism and leads to second-order field equations \cite{SJ:crossroads,Jansen:palatini,Pons:palatini}. 

In this letter we will deepen this analogy with general relativity. Notions such as the Riemann tensor, the metricity condition and the Levi-Civita connection have a Galilean counterpart. Our results place Galileons on a par with Lovelock gravity. The Palatini formalism provides an elegant derivation and proof of the second-order nature of both theories. Finally, this improved understanding might allow for a similar construction in other domains as well.

\section{General Relativity}

General relativity is based on diffeomorphism invariance. This dictates the form of the covariant field strength of the metric, the Riemann tensor. Since the Riemann tensor contains second derivatives of the metric, a generic theory depending on it will have field equations that are higher order in derivatives than two. The Einstein-Hilbert term is special as it yields only second-order field equations. F.e, all other $f(R)$-gravity theories have higher-order terms leading to an additional scalar degree of freedom, in addition to those of a massless graviton. 

In addition to the Einstein-Hilbert term, there is a larger subset of general relativity theories that have up to second order field equations. This set is referred to as Lovelock gravity \cite{Lovelock}, and its basic building blocks are the Lovelock invariants $\mathcal{L}_0 =\sqrt{-g}$ and
\begin{align} \label{Lovelock}
\mathcal{L}_n = \sqrt{-g} \delta^{\m_1\ldots\m_{2n}}_{\n_1\ldots\n_{2n}} \prod_{i=1}^n R_{\mu_{2i-1} \mu_{2i}}{}^{\nu_{2i-1} \nu_{2i}} \,,
\end{align}
with $n=1,2,\ldots$. In addition to the cosmological constant and the Einstein-Hilbert term, this includes e.g. at quadratic order the Gauss-Bonnet term.
The most general Lovelock Lagrangian in $D$-dimensions is a linear combination of the Lovelock invariants with $n\leq [(D-1)/2]$, since the higher order invariants are non-dynamical in $D$-dimensions.

The set of Lovelock theories can be derived in an elegant and insightful manner by means of a first-order formulation.
In this so-called Palatini formalism \cite{Palatini}, one assumes both geometric objects of general relativity, the metric $g_{\mu\nu}$ and the connection $\Gamma^{\r}{}_{\m\n}$, to be independent - in contrast to the usual metric formalism in which one takes the Levi-Civita connection. Generically, evaluating a theory in the metric formalism or in the Palatini formalism yields entirely different dynamics. Only for Lovelock theories there is a clear relation between the two formalisms: the dynamics of the metric formalism is contained in that of the Palatini formalism. One can see this by starting from a generic theory of the form 
\begin{align}
\mathcal{L} = \mathcal{L}(g_{\mu\nu}, 
R_{\m\n\r}{}^\s = 2\partial_{[\n} \G^\s{}_{\m]\r} + 2\G^\l{}_{[\m|\r} \G^\s{}_{\l|\n]} ) \,.
\end{align}
In the metric formalism the equation of motion is
\begin{align}
\frac{\delta \mathcal{L}}{\delta g_{\mu \nu}} = \frac{\partial \cL}{\partial g_{\m\n}} + 2\nabla_{(\r} \nabla_{\s)} \big( g^{\mu \l} \frac{\partial \cL}{\partial R_{\n\r\s}{}^\l} \big) \,,
\label{metriceom}
\end{align}
whilst in the Palatini formalism the equations of motion are
\begin{align}
\frac{\delta \mathcal{L}}{\delta g_{\mu \nu}} &= \frac{\partial \cL}{\partial g_{\m\n}} = 0 \,,\label{palatinimetriceom} \\
\frac{\delta \mathcal{L}}{\delta \Gamma^\rho{}_{\m \n}} &=  \nabla_\s \Big( \frac{\partial \cL}{\partial R_{\s(\m\n)}{}^\r} \Big) = 0 \,.
\label{palatiniconnectioneom}
\end{align}
(Here we assumed the absence of torsion, see \cite{Pons:palatini} for a more general treatment.) For Lagrangians that are linear in the Riemann tensor, i.e.~the Einstein-Hilbert term, \eqref{palatiniconnectioneom} necessarily implies the {metricity condition}
\begin{align}
\nabla_{\r} g_{\m\n} = 0 \,,
\end{align}
which determines the connection to be the Levi-Civita connection. Upon using this unique solution, \eqref{metriceom} and \eqref{palatinimetriceom} are seen to be identical. If higher powers of the Riemann tensor are present, the Levi-Civita connection is generically no longer a solution to \eqref{palatiniconnectioneom}. However, following \cite{SJ:crossroads}, we will require it to be a solution to higher-order terms in the Lagrangian as well, thus restricting the set of admissible Lagrangians. 

We now assume our Lagrangian admits an expansion in terms of the Riemann tensor and whose coefficients only depend on the metric. {Since} the latter are covariantly constant due to the {metricity condition}, we can apply the chain rule to obtain
\bea
\frac{\partial^2 \cL}{\partial R_{\s(\m\n)}{}^\r \partial R_{\a\b\g}{}^\d} \nabla_\s R_{\a\b\g}{}^\d = 0 \,.
\eea
By using the Bianchi identity $\nabla_{[\s} R_{\a\b]\g}{}^\d = 0$, this can be translated in a condition on the Lagrangian:
\bea
\frac{\partial^2 \cL}{\partial R_{(\s|(\m\n)}{}^\r \partial R_{\a)\b\g}{}^\d} = 0 \,.
\label{MetricPalatiniCondition}
\eea
This condition uniquely singles out the Lovelock invariants $\mathcal{L}_n$. In addition, the second term of \eqref{metriceom} vanishes on account of the Lovelock condition \eqref{MetricPalatiniCondition} and one concludes that \eqref{metriceom} and \eqref{palatinimetriceom} are identical upon restricting to the Levi-Civita connection.
Moreover, this demonstrates the second-order nature of the field equations for these theories: as the Lagrangian only includes the metric and the Riemann tensor, its variation with respect to explicit metrics can only contain up to second-order derivatives. 

\section{Galilean Theories}
A generic Galilean theory will have a field equation higher order in derivatives than two. The subset of such theories that have up to second order equations of motion, are the Galileons of \cite{Rattazzi}, whose basic building blocks can be written as \cite{Kurt} (see also \cite{Wenliang})
\begin{align}
\mathcal{L}_n = \phi^{n+1}\delta^{\m_1\ldots\m_n}_{\n_1\ldots\n_n}\prod_{i=1}^{n} S_{\m_i}{}^{\n_i},\quad  n=1,2,\ldots 
\label{Galileons}
\end{align}
where we introduced
\begin{align} \label{S}
S_{\mu \nu} = \phi^{-1} \partial_\mu \partial_\nu \phi \,.
\end{align}
The most general Galileon in $D$-dimensions is a linear combination of the invariants with $n \leq D$, since all higher-order ones vanish.

By comparing \eqref{Lovelock} and \eqref{Galileons}, one is tempted to view $S_{\m\n}$ as a Galilean covariant field strength for the scalar. To see this, consider a scalar field $\phi$ whose transformation we write as
\bea \label{scalarvariation}
\d \phi &=& \L (x) = \frac{\L (x)}{\phi} \phi \,,
\eea
where $\L (x)$ is a space-time dependent parameter. In order to construct covariant quantities, we introduce a gauge field that transforms as
\bea
\d \G_\m &=& \partial_\m \Big( \frac{\L (x)}{\phi} \Big) \,.
\eea
The covariant derivative is then given by
\bea
\cD_\m \phi &=& (\partial_\m - \G_\m ) \phi  \,.
\eea
Moreover, we define a field strength for $\G_\m$ as
\bea
S_{\m\n} &=& \partial_\m \G_\n + \alpha \G_\m \G_\n \,, \label{field-strength}
\eea
whose transformation is given by
\begin{align} \label{transformation-S}
 & \d S_{\m\n}  =  - \frac{\L}{\phi}  S_{(\m\n)} + \frac1{\phi} \partial_\m \partial_\n \L - \frac1\L \cD_{(\mu} \Big( \frac{\L^2}{\phi^2} \cD_{\nu)} \phi \Big)  \nn\\ 
& + \frac1\phi (1-\a)\G_{(\m} \Big(  \G_{\n)} \L - 2 \partial_{\n)} \L   + \frac{2\L}{\phi}  \cD_{\n)} \phi   \Big) \,.
\end{align}
In order for this to be covariant one usually takes the field strength to be the anti-symmetric combination. The transformation then vanishes identically and the field strength is invariant. This leads to the usual set of gauge theories. In contrast, we will take the symmetric projection, which forces us to impose the condition $\L = a + b_\mu x^\mu$ and set $\alpha = 1$ in order to eliminate the non-covariant terms. This implies that we have Galilean symmetry acting on our scalar field. 

Note however that in this case the field strength is not invariant, but merely covariant. Also note that it is due to the symmetry of the field strength that we can write down its second term (which is reminiscent of a non-Abelian term but absent in the case of a single gauge field). We stress that this form, which is suggestively similar to the Riemann tensor, is dictated by symmetry; apart from gauge theories, the unique alternative are Galilean theories.

To make contact with definition \eqref{S}, we need to impose a condition expressing $\Gamma_\m$ in terms of the scalar. The only covariant option is the vanishing of the covariant derivative:
\begin{align} \label{Gamma}
\cD_\m \phi = 0 \,, \quad \Leftrightarrow \quad \Gamma_\mu = \phi^{-1} \partial_\mu \phi \,,
\end{align}
which allows us to solve the connection as a composite field (note that it can always be set equal to zero at any point by means of a Galilean transformation). Inserting this expression into \eqref{field-strength} (with $\alpha=1$) precisely yields \eqref{S}. \footnote{More generally $\cD_{\r_1}\dots \cD_{\r_n}S_{\m\n}$, upon using \eqref{Gamma}, reduces to $\phi^{-1}\partial_{\r_1}\dots \partial_{\r_n}\partial_\m\partial_\n \phi$. }

Now, starting from a Galilean theory written in terms of $\phi$ and $S_{\m\n}[\Gamma_\r]$ \footnote{We suspect that any Galilean theory $\mathcal{L}(\phi,\partial\phi,\partial\partial\phi,...)$ can be rewritten as $\mathcal{L}'(\phi,\partial\partial\phi,...)$ by adding suitable total derivatives, and hence that any Galilean theory can be written in terms of $\phi$, $S_{\m\n}$ and covariant derivatives of $S_{\m\n}$ only.}, two different formalisms naturally present themselves. Firstly, one can take the 'metric' approach by demanding the vanishing of the covariant derivative and fully expressing the theory in terms of the scalar, resulting in the usual formulation. Secondly, one can take a Palatini approach and view $\Gamma_\m$ as an independent field. As in the case of general relativity, the two formalisms generically yield different dynamics. However, precisely for Galileons do the two formalisms have a clear relation: the dynamics of the metric formalism is contained in that of the Palatini formalism.

To see this consider a general Lagrangian
\begin{align}
\mathcal{L} = \mathcal{L} ( \phi , S_{\mu \nu} = \partial_{(\mu} \G_{\nu)} + \G_\mu \G_\nu) \,,
\end{align}
whose field equation in the metric formalism is
\begin{align} \label{EOM-G}
\frac{\delta \mathcal{L}}{\delta \phi} =  \frac{\partial \mathcal{L}}{\partial \phi} + (\partial_\mu +  \Gamma_\mu) (\partial_\nu - \Gamma_\n)  \big( \frac{1}{\phi} \frac{\partial \mathcal{L}}{\partial S_{\mu \nu}} \big) \,.
\end{align}
and whose field equations in the Palatini formalism are
\begin{align} \label{EOM-Palatini}
\frac{\delta \mathcal{L}}{\delta \phi} &=  \frac{\partial \mathcal{L}}{\partial \phi} = 0 \\
\frac{\delta \mathcal{L}}{\delta \Gamma_\m} &= (\partial_\n - 2 \Gamma_\n) (\frac{ \partial\mathcal{L}}{\partial S_{\m\n}}) =0\,.
\label{Palatini-G}
\end{align}
For the Galilean theory linear in the field strength, i.e. $\phi^2 S^{\m}_\m$, \eqref{Palatini-G} implies the vanishing of the covariant derivative and the two formalisms are seen to be equivalent. However, for higher order terms the vanishing of the covariant derivative is no longer consistent with \eqref{Palatini-G}. Following the discussion in the previous section, we will require \eqref{Gamma} to be a solution to the higher order terms as well, thus restricting the set of admissible Lagrangians.

Assuming a Taylor expansion in $S_{\m\n}$ with $\phi$ dependent coefficients and using \eqref{Gamma} at the level of the equations of motion, one finds firstly that the Lagrangian must have weight +1, and secondly
\begin{align} \label{condition}
\frac{ \partial^2\mathcal{L}}{\partial S_{\mu\nu} \partial S_{\r\s}} \cD_\mu S_{\r\s} =0 \,.
\end{align}
Note that the field strength transforms as $\phi^{-1}$, which we will refer to as weight $-1$ (see first term in \eqref{transformation-S}). Its covariant derivative is therefore defined as $\cD_\m = \partial_\m + \Gamma_\m$ and satisfies the Bianchi identity
	\begin{align}
	\cD _{[\r} S_{\m]\n} = 0 \,.
	\end{align}
Using this identity, equation \eqref{condition} translates into the following condition on the Lagrangian
\begin{align}\label{condition2}
\frac{ \partial^2\mathcal{L}}{\partial S_{\mu(\nu} \partial S_{\r)\s}} = 0
\end{align}

These conditions, namely weight +1 and \eqref{condition2}, ensure that the second term of \eqref{EOM-G}  vanishes. Therefore the scalar field equations \eqref{EOM-G} and \eqref{EOM-Palatini}  are identical upon restricting to solution \eqref{Gamma} of \eqref{Palatini-G}. Similar to general relativity, this demonstrates the second-order nature of the field equations for these theories: the Lagrangian only depends on $\phi$ and $S_{\m\n}$ implying up to second-order derivatives in the variation with respect to the explicit $\phi$'s. 
	
Finally, the conditions uniquely single out the Galileons. To see this we note that any Lagrangian satisfying the conditions yields a purely second order scalar field equation, and is hence contained in the Galileons. Conversely, any Galileon has weight +1 and satisfies \eqref{condition2}.

\section{Generalized Galileons}

Remarkably, one can extend the first-order formalism beyond theories that are restricted by symmetries, as in the above. The set of theories can be extended to the generalized Galileons, which have an arbitrary function in front of the Galileon invariants, and still have up to second-order field equations \cite{Deffayet:k-essence}. However, while the most general set allows for dependence on both $\phi$ as well as $X \equiv (\partial \phi)^2$, we restrict to functions $f(\phi)$ only.

We will again start from a scalar field transforming as \eqref{scalarvariation}. However, in this case we introduce a gauge field with transformation
\bea
\d \G_\m &=& \partial_\m \Big( \frac{\L (x)}{f(\phi)} \Big) \,.
\eea
The covariant derivatives and field strength read
\begin{align}
\cD_\m \phi & =  \partial_\m \phi  - \G_\m f \,, \notag \\
\cD_\m f & =  ( \partial_\m  - \G_\m f') f \,, \notag \\
S_{\m\n} & =  \partial_{(\mu}\Gamma_{\n)} + f'\Gamma_\m \Gamma_\n  \,,
\end{align}
where $f^\prime (\f) \equiv \partial f(\f) / \partial \f$. The vanishing of the covariant derivative in this case yields
\begin{align} \label{fconnection}
\Gamma_\m &= f^{-1}\partial_\m\phi \,, \quad \Rightarrow \quad
S_{\mu \nu} = f^{-1} \partial_\mu \partial_\nu \phi \,.
\end{align}

Starting from a Lagrangian with arbitrary $\phi$ and $S_{\mu \nu}$ dependence, the equation of motion for the connection is:
\begin{align}
( \partial_\n - 2 f' \Gamma_\n) (\frac{\partial\mathcal{L}}{\partial S_{\m\n}} ) = 0 \,.
\end{align}
A similar analysis as in the previous section reveals that the only Lagrangians whose connection equation has the solution \eqref{fconnection} are of the form
\begin{align}
\mathcal{L}_n = f(\phi)^{n+1}\delta^{\m_1\ldots\m_n}_{\n_1\ldots\n_n}\prod_{i=1}^{n} S_{\m_i}{}^{\n_i},\quad  n=1,2,\ldots 
\end{align}
which are precisely the generalized Galileons whose free function is restricted to be $\phi$ dependent only.

Again the equivalence of the scalar equations of motion follow automatically. Note however that the scalar field equation is no longer obtained by solely varying with respect to the explicit $\phi$'s: it picks up contributions from the variation of $S_{\m\n}$.

This first-order formalism therefore leads to the generalized Galileons with arbitrary functional dependence on the scalar field (but not its first derivatives). In the case of a constant function, the Lagrangian is invariant under the Galilean transformation; however, all these terms are total derivatives and hence not dynamical.
The case of a linear function leads to the actual Galileons; the Lagrangian transforms with a total derivative. For quadratic and higher-order expressions, the resulting Lagrangians are not invariant. Nevertheless, it provides one with an understanding of why these theories lead to second-order field equations: the consistency and equivalence of the Palatini formalism only requires the Lagrangian to be covariant, and not invariant, under the underlying symmetry. 

\section{Conclusions}
\begin{table}[b!]
\begin{tabular}{|l||c|c|}
\hline
& \it General relativity & \it Galilean theories \\ \hline \hline
\it Field & Metric $g_{\m\n}$ & Scalar $\phi$ \\ \hline
\it Symmetry & Diffeomorphism & $\phi \rightarrow \phi + c + v_\mu x^\mu$ \\ \hline
\it Connection & $\Gamma^\r{}_{\m\n}$ & $\Gamma_\m$ \\ \hline
\it Field strength & $R_{\mu\nu \rho}{}^{\sigma} $ &  $S_{\mu\nu} \equiv  \partial_{(\mu} \Gamma_{\nu)} + \Gamma_\mu \Gamma_\nu$ \\ \hline
\it Bianchi & $\nabla_{[\l} R_{\m\n]\r}{}^\s = 0$ & $\cD _{[\r} S_{\m]\n} = 0$   \\ \hline
\it Condition & $\nabla_\rho g_{\mu\nu} = 0$ & $\cD_{\mu} \phi = 0$ \\ \hline
\it 2$^{\rm nd}$-order & Lovelock & Galileons \\ \hline 
\end{tabular}
\caption{\it Dictionary of concepts in GR and Galilean theories.}
\label{table}
\vspace{-0.5cm}
\end{table}
To conclude, it seems worthwhile to summarize the similarities (and differences) between the first-order formalisms for the metric and the scalar. We have summarized the analogous concepts in Table \ref{table} to highlight the striking similarity. 

One difference concerns the transformation of the field strength in the two formalism. While the Riemann tensor transforms identically in the first- and second-order formulation, this is not the case for $S_{\mu \nu}$: in addition to the covariant weight $-1$ piece, there are additional terms in the last line of \eqref{transformation-S}. However, since these all vanish upon imposing the vanishing of the covariant derivative, this difference is immaterial. 
Similarly, one could worry about the lack of covariance of the field equations \eqref{EOM-G} and \eqref{Palatini-G}. Indeed generic Lagrangians will not be invariant (up to a total derivative) under Galilean transformations, resulting in non-covariant field equations. However, these terms vanish exactly for the set of Galileons.

Galileons can be constructed as Wess-Zumino terms in non-linear realizations of the appropriate coset space \cite{Brugues, Trodden}. This construction does not entail a one-to-one correspondence between broken generators and Goldstone bosons -- while $\phi$ is the Goldstone boson of broken translational invariance in the fifth dimension, there is a so-called inverse Higgs mechanism \cite{Ogievetsky} at work that expresses the Goldstone bosons of the other four broken generators in terms of the scalar field. This is reminiscent of our covariant derivative condition \eqref{Gamma}. Similarly, one can construct conformal and DBI Galileons \cite{DGP, deRham-brane, Trodden} by starting from different cosets; do these also admit an underlying Palatini formalism? More generally, does this formalism extend to generalized Galileons with arbitrary dependence on the first derivative of the scalar as well  (including DBI and conformal)? It is unclear how to extend the formalism to cope with such terms in general; e.g.~there is an ambiguity in how to express $X$ in terms of $\Gamma_\m$ and $\partial_\m\phi$.

In addition, one can couple the Galileons to gravity. The resulting Lagrangian reads \cite{Deffayet:covariant,Deffayet:generalized}
\begin{align}
\mathcal{L}_{n} =&  \sqrt{-g} \phi \sum_{p=0}^{[n/2]}  (-\frac{1}{8})^p \frac{n!}{(n-2p)!p!p!}   \delta^{\m_1\ldots\m_n}_{\n_1\ldots\n_n} \cdot \notag \\
 & \cdot \prod_{i=1}^{p} X R_{\m_{2i-1} \m_{2i}}{}^{\n_{2i-1} \n_{2i}} \cdot \prod_{j=2p+1}^{n} \phi S_{\m_j}{}^{\n_j} \,.
\end{align}
Note that these terms interpolate between the Galileon \eqref{Galileons} and Lovelock \eqref{Lovelock} structures; however, the latter are polynomial in $X R$ rather than $R$ itself. Moreover, this theory can be generalized by including an arbitrary function $f_n( \phi,X )$ for every $n$ \cite{Deffayet:k-essence}, without losing the second-order nature of the field equations. In four dimensions this set of theories is actually identical to the much older Horndeski result \cite{Horndeski,FabFour}. Finally, it has been argued that a further generalization allows for more general functional dependence without introducing additional propagating degrees of freedom \cite{Zuma:transforming,GLPV:beyond,Deffayet:counting, Langlois}.

It is natural to wonder whether such scalar-tensor theories also allow for a first-order formalism, and to interpret the resulting geometric structures. A possible starting point for such an extension might be found in the dimensional reduction of higher-dimensional Lovelock theories, giving rise to Horndeski-type theories \cite{KKreduction} (or brane-world scenarios \cite{DGP, deRham-brane}). This connection to higher-dimensional gravitational theories might point the way to a first-order formalism for scalar-tensor theories; moreover, it could provide a geometric interpretation for the various structures uncovered in this letter.\\

\section*{Acknowledgments}
We thank Joaquim Gomis for stimulating discussions. RK acknowledges the Dutch funding agency `Foundation for Fundamental Research on Matter' (FOM) for financial support.

\providecommand{\href}[2]{#2}\begingroup\raggedright\endgroup


\begin{thebibliography}{10}

\bibitem{MO}
M.~Ostrogradsky. Mem. Ac. St. Petersbourg {\bf VI 4} (1850) 385.

\bibitem{Rattazzi}
A.~Nicolis, R.~Rattazzi and E.~Trincherini,  {\em {The Galileon as a local
		modification of gravity}}, Phys. Rev. {\bf D79} (2009) 064036
[\href{http://www.arXiv.org/abs/0811.2197}{{\tt 0811.2197}}].

\bibitem{Deffayet:generalized}
C.~Deffayet, S.~Deser and G.~Esposito-Farese,  {\em {Generalized Galileons: All
		scalar models whose curved background extensions maintain second-order field
		equations and stress-tensors}}, Phys. Rev. {\bf D80} (2009) 064015
[\href{http://www.arXiv.org/abs/0906.1967}{{\tt 0906.1967}}].

\bibitem{Fairlie}
D.~B. Fairlie, J.~Govaerts and A.~Morozov,  {\em {Universal field equations
		with covariant solutions}}, Nucl. Phys. {\bf B373} (1992) 214--232
[\href{http://www.arXiv.org/abs/hep-th/9110022}{{\tt hep-th/9110022}}].

\bibitem{DGP}
G.~R. Dvali, G.~Gabadadze and M.~Porrati,  {\em {4-D gravity on a brane in 5-D
		Minkowski space}}, Phys. Lett. {\bf B485} (2000) 208--214
[\href{http://www.arXiv.org/abs/hep-th/0005016}{{\tt hep-th/0005016}}].

\bibitem{Deffayet:2000uy}
C.~Deffayet,  {\em Cosmology on a brane in Minkowski bulk}, Phys. Lett. B {\bf
	502} (2001) 199--208 [\href{http://www.arXiv.org/abs/hep-th/0010186}{{\tt
		hep-th/0010186}}].

\bibitem{Deffayet:2001pu}
C.~Deffayet, G.~Dvali and G.~Gabadadze,  {\em {Accelerated universe from gravity
	leaking to extra dimensions}}, Phys. Rev. D {\bf 65} (2002)
[\href{http://www.arXiv.org/abs/astro-ph/0105068}{{\tt astro-ph/0105068}}].

\bibitem{NR1}
M.~A. Luty, M.~Porrati and R.~Rattazzi,  {\em {Strong interactions and
		stability in the DGP model}}, JHEP {\bf 09} (2003) 029
[\href{http://www.arXiv.org/abs/hep-th/0303116}{{\tt hep-th/0303116}}].

\bibitem{NR2}
K.~Hinterbichler, M.~Trodden and D.~Wesley,  {\em {Multi-field galileons and
		higher co-dimension branes}}, Phys. Rev. {\bf D82} (2010) 124018
[\href{http://www.arXiv.org/abs/1008.1305}{{\tt 1008.1305}}].

\bibitem{GalInf}
C.~Burrage, C.~de~Rham, D.~Seery and A.~J. Tolley,  {\em {Galileon inflation}},
JCAP {\bf 1101} (2011) 014
[\href{http://www.arXiv.org/abs/1009.2497}{{\tt 1009.2497}}].

\bibitem{NEC}
K.~Hinterbichler, A.~Joyce, J.~Khoury and G.~E.~J. Miller,  {\em
	{Dirac-Born-Infeld Genesis: An Improved Violation of the Null Energy
		Condition}}, Phys. Rev. Lett. {\bf 110} (2013), no.~24, 241303
[\href{http://www.arXiv.org/abs/1212.3607}{{\tt 1212.3607}}].

\bibitem{FabFour}
C.~Charmousis, E.~J. Copeland, A.~Padilla and P.~M. Saffin,  {\em {General
		second order scalar-tensor theory, self tuning, and the Fab Four}}, Phys.
Rev. Lett. {\bf 108} (2012) 051101
[\href{http://www.arXiv.org/abs/1106.2000}{{\tt 1106.2000}}].

\bibitem{Gomis:2000}
J.~Gomis and H.~Ooguri,  {\em {Nonrelativistic closed string theory}}, J. Math.
Phys. {\bf 42} (2001) 3127--3151
[\href{http://www.arXiv.org/abs/hep-th/0009181}{{\tt hep-th/0009181}}].

\bibitem{Gomis:2004}
J.~Gomis, K.~Kamimura and P.~K. Townsend,  {\em {Non-relativistic
		superbranes}}, JHEP {\bf 11} (2004) 051
[\href{http://www.arXiv.org/abs/hep-th/0409219}{{\tt hep-th/0409219}}].

\bibitem{SJ:crossroads}
Q.~Exirifard and M.~M. Sheikh-Jabbari,  {\em {Lovelock gravity at the
		crossroads of Palatini and metric formulations}}, Phys. Lett. {\bf B661}
(2008) 158--161
[\href{http://www.arXiv.org/abs/0705.1879}{{\tt 0705.1879}}].

\bibitem{Jansen:palatini}
M.~Borunda, B.~Janssen and M.~Bastero-Gil,  {\em {Palatini versus metric
		formulation in higher curvature gravity}}, JCAP {\bf 0811} (2008) 008
[\href{http://www.arXiv.org/abs/0804.4440}{{\tt 0804.4440}}].

\bibitem{Pons:palatini}
N.~Dadhich and J.~M. Pons,  {\em {Consistent Levi Civita truncation uniquely
		characterizes the Lovelock Lagrangians}}, Phys. Lett. {\bf B705} (2011)
139--142
[\href{http://www.arXiv.org/abs/1012.1692}{{\tt 1012.1692}}].

\bibitem{Lovelock}
D.~Lovelock,  {\em {The Einstein tensor and its generalizations}}, J. Math.
Phys. {\bf 12} (1971)
498--501.

\bibitem{Palatini}
A.~Palatini,  {\em {Deduzione invariantiva delle equazioni gravitazionali dal
		principio di Hamilton}}, Rend. Circ. Mat. Palermo {\bf 43} (1919) 203.

\bibitem{Kurt}
K.~Hinterbichler and A.~Joyce,  {\em {Hidden symmetry of the Galileon}}, Phys.
Rev. {\bf D92} (2015), no.~2, 023503
[\href{http://www.arXiv.org/abs/1501.07600}{{\tt 1501.07600}}].


\bibitem{Wenliang}
W.~Li, {\em{Unifying Ghost-Free Lorentz-Invariant Lagrangians}},[\href{http://www.arXiv.org/abs/1510.05496}{{\tt 1510.05496}}].


\bibitem{Deffayet:k-essence}
C.~Deffayet, X.~Gao, D.~A. Steer and G.~Zahariade,  {\em {From k-essence to
		generalised Galileons}}, Phys. Rev. {\bf D84} (2011) 064039
[\href{http://www.arXiv.org/abs/1103.3260}{{\tt 1103.3260}}].

\bibitem{Brugues}
J.~Brugues, J.~Gomis and K.~Kamimura,  {\em {Newton-Hooke algebras,
		non-relativistic branes and generalized pp-wave metrics}}, Phys. Rev. {\bf
	D73} (2006) 085011
[\href{http://www.arXiv.org/abs/hep-th/0603023}{{\tt hep-th/0603023}}].

\bibitem{Trodden}
G.~Goon, K.~Hinterbichler, A.~Joyce and M.~Trodden,  {\em {Galileons as
		Wess-Zumino Terms}}, JHEP {\bf 06} (2012) 004
[\href{http://www.arXiv.org/abs/1203.3191}{{\tt 1203.3191}}].

\bibitem{Ogievetsky}
E.~A. Ivanov and V.~I. Ogievetsky,  {\em {The Inverse Higgs Phenomenon in
		Nonlinear Realizations}}, Teor. Mat. Fiz. {\bf 25} (1975)
164--177.

\bibitem{deRham-brane}
C.~de~Rham and A.~J. Tolley,  {\em {DBI and the Galileon reunited}}, JCAP {\bf
	1005} (2010) 015
[\href{http://www.arXiv.org/abs/1003.5917}{{\tt 1003.5917}}].

\bibitem{Deffayet:covariant}
C.~Deffayet, G.~Esposito-Farese and A.~Vikman,  {\em {Covariant Galileon}},
Phys. Rev. {\bf D79} (2009) 084003
[\href{http://www.arXiv.org/abs/0901.1314}{{\tt 0901.1314}}].

\bibitem{Horndeski}
G.~W. Horndeski,  {\em {Second-order scalar-tensor field equations in a
		four-dimensional space}}, Int. J. Theor. Phys. {\bf 10} (1974)
363--384.

\bibitem{Zuma:transforming}
M.~Zumalacárregui and J.~García-Bellido,  {\em {Transforming gravity: from
		derivative couplings to matter to second-order scalar-tensor theories beyond
		the Horndeski Lagrangian}}, Phys. Rev. {\bf D89} (2014) 064046
[\href{http://www.arXiv.org/abs/1308.4685}{{\tt 1308.4685}}].

\bibitem{GLPV:beyond}
J.~Gleyzes, D.~Langlois, F.~Piazza and F.~Vernizzi,  {\em {Exploring
		gravitational theories beyond Horndeski}}, JCAP {\bf 1502} (2015) 018
[\href{http://www.arXiv.org/abs/1408.1952}{{\tt 1408.1952}}].

\bibitem{Deffayet:counting}
C.~Deffayet, G.~Esposito-Farese and D.~A. Steer,  {\em {Counting the degrees of
		freedom of generalized Galileons}},
\href{http://www.arXiv.org/abs/1506.01974}{{\tt 1506.01974}}.

\bibitem{Langlois}
D.~Langlois and K.~Noui,  {\em {Degenerate higher derivative theories beyond
		Horndeski: evading the Ostrogradski instability}},
\href{http://www.arXiv.org/abs/1510.06930}{{\tt 1510.06930}}.

\bibitem{KKreduction}
K.~Van~Acoleyen and J.~Van~Doorsselaere,  {\em {Galileons from Lovelock
		actions}}, Phys. Rev. {\bf D83} (2011) 084025
[\href{http://www.arXiv.org/abs/1102.0487}{{\tt 1102.0487}}].




\end{thebibliography}
\end{document}